\documentclass[a4paper,aps,twocolumn,pra,showpacs,floatfix]{revtex4-1}

\usepackage{graphicx}
\usepackage{amsmath}
\usepackage{amstext}
\usepackage{graphicx}
\def\S{\mathcal{S}}
\def\H#1{\mathcal{B}(\mathcal{H_{\text{#1}}})}

\def\tr{\mathop{\mbox{tr}}}
\begin{document}
\title{Process optimized quantum cloners via semidefinite programming}

\author{M\'aty\'as Koniorczyk
\affiliation{Institute of Mathematics, University of P\'ecs, H-7624  P\'ecs, Ifj\'us\'ag \'utja 6, Hungary}
}

\author{L\'\i via Dani}
\affiliation{Institute of Physics, University of P\'ecs, H-7624  P\'ecs, Ifj\'us\'ag \'utja 6, Hungary}

\author{Vladim\'\i r Bu\v zek}
\affiliation{Research Center for Quantum Information, Institute of Physics, Slovak Academy of Sciences}
\affiliation{Dubravsk\'a cesta 9, 84511 Bratislava, Slovakia}

\date{September 12. 2012.}
\begin{abstract}
  We apply semidefinite programming for designing 1 to 2 symmetric
  qubit quantum cloners. These are optimized for the average fidelity
  of their joint output state with respect to a product of multiple
  originals. We design 1 to 2 quantum bit cloners using the numerical
  method for finding completely positive maps approximating a
  nonphysical one optimally. We discuss the properties of the
  so-designed cloners.
\end{abstract}
\maketitle

\section{Introduction}
\label{sec:intro}

A quantum process describes what can happen to a physical system when
its state changes. Mathematically it is a mapping from the set of
the initial states of the system to that of its final states. Quantum
mechanics provides well-defined limitations to the process to be
physically realistic. Apart from having to be linear, positive and
trace preserving, it should be completely positive, too. This is an
important constraint in the design of any quantum information
processing apparatus. 

Quantum cloning is a celebrated counterexample of a physically
realistic process, producing identical copies of a physical system in
a given unknown quantum state. This mapping is not even linear, hence
it is unfeasible physically. Of course, if something is unfeasible, it
can still be approximated, as described first in
Ref.~\cite{PhysRevA.54.1844}. Since that, the topic of cloning achieved
a broad coverage in the literature (see Ref.~\cite{cloning} for a review),
including the calculation of achievable fidelities and several designs
of particular schemes for cloning. These designs of optimal cloners
are based on some intuitive physical ideas, which makes them well
understandable and gives a hint for laboratory interpretation.

It is also known that the methods of semidefinite programming in
operations research makes it possible to find completely positive maps
ideally approximating nonphysical ones. In Ref.~\cite{koenraad}, where
this idea was introduced, an example is given which is a universal
shifter~\cite{PhysRevA.63.032304,PhysRevA.64.062310,PhysRevA.64.032301},
a nonlinear operation on a single qubit. Besides of the valuable
analytical techniques, Ref~\cite{koenraad}. gives a complete numerical
recipe to design arbitrary operations.

In the present work we utilize this numerical technique in order to
analyze the design of 1 to 2 quantum bit cloners. It turns out that
partly due to the nature of the optimization method, the so designed
cloner is not optimized with respect to the fidelity of the clones to
the input state, but the average fidelity of the whole (two-qubit
output) with respect to a tensor product of two copies of the
original. 
% This is an approach interesting \emph{per se}, even
% independently of the actual numerical or analytical optimization
% method. We discuss in detail the relation of the so-designed cloners
% to those designed for optimal clone fidelity. The dependence of
% the cloner operation on the initial state of an ancilla, as well as
% the effect of restricting the possible input states is discussed.
For universal symmetric quantum cloners, that is, those which work for
an arbitrary input qubit state, it was shown by Werner~\cite{Werner}
and Keyl and Werner~\cite{KeylWerner} in a much more general context
that optimizing the fidelity of the two qubit output state is
equivalent to that of optimizing the fidelity of the two cloners. For
phase-covariant cloners, that is, for which the input state is
restricted to those on a main circle of the Bloch sphere, the two
aspects are not equivalent. The design of the phase covariant cloners
for qubits and qutrits is studied in detail in the paper of D'Ariano
and Macchiavello in Ref.~\cite{DAriano}. Even though the existence and
some properties are covered by these papers partly, we the use
different, numerical approach, an application of the general method in
Ref~\cite{koenraad}. This provides us with easily applicable mappings
and demonstrates the power of semidefinite programming in this field. It
is also interesting to compare the maps we have found with the
universal covariant quantum cloner (UCQC) circuit designed by
Braunstein et al.~\cite{PhysRevA.63.052313}.

This paper is organized as follows. In Section~\ref{sec:method} we
review the method for finding optimal CP approximations of nonphysical
maps via semidefinite programming, and discuss its application to
quantum cloning, introducing the notion of process optimized
cloners. In Section~\ref{sec:cloners} we present and discuss our
particular results. In Section~\ref{sec:conclusion} conclusions are
drawn and an outlook is provided.

\section{Design method}
\label{sec:method}

In this Section we first briefly summarize the method for designing CP
maps via semidefinite programming, which optimally approximate an
unphysical map. This is entirely described in detail in
Ref.~\cite{koenraad}, we just repeat the main ideas here for sake of
self-consistency. In the second subsection, the application of this
method to the design of quantum cloners is discussed, which leads us
to the concept of process-optimized cloners.

\subsection{Optimizing CP maps via semidefinite programming}

Consider a quantum operation $\S: \H{in} \mapsto \H{out}$, mapping
states in the Hilbert space $\mathcal{H}_\text{in}$ to those of
$\mathcal{H}_\text{out}$, $\mathcal{B}$ denoting the set of density
operators on the given Hilbert space. In order to be physically
realistic, the mapping $\S$ has to be a linear, completely positive
(CP) trace preserving one. In some cases, however, one might want to
at least approximately realize processes which are non-CP, or not even
linear. Examples include the quantum shifter discussed in
Refs.~\cite{PhysRevA.63.032304,PhysRevA.64.062310,PhysRevA.64.032301}
and quantum cloning studied here, etc.

So we consider an ideal process $\S_\text{id}$ which is arbitrary, and
we seek a realistic (linear, trace-preserving, CP) map $\S$ which
approximates $\S_\text{id}$ \emph{to the highest extent}. This latter
should be quantified somehow. In order to do so, consider an
\emph{input set} $T\subset \H{in}$, containing the states for which we
would like our approximate process to be optimal. (This possible
restriction might be of some use, it enables us, for instance, to
consider non-universal quantum cloners in this framework.) We assume
that it is possible to integrate over $T$ according to a suitable
measure, this will be denoted by $\int_T dT$.  The quantity to be
optimized will be
\begin{equation}
  \label{eq:processfidelity}
  \mathcal{F}=\int_TdT \left(\tr 
  \sqrt{\sqrt{\S_{\text{id}}(\varrho)}
    \S(\varrho)
    \sqrt{\S_{\text{id}}(\varrho)}}\right)^2,\quad
  (\varrho\in T),
\end{equation}
the fidelity of the output state of the realistic process with respect
to that of the desired ideal process, averaged over the considered
input states. We shall term this as \emph{process fidelity} in what
follows. In the special case when the set
$\{\S_{\text{id}}(\varrho)|\varrho \in T\}$ contains pure states only,
that is, we expect the ideal process to map the states of the target
space to pure states only,
the fidelity in Eq.~\eqref{eq:processfidelity} simplifies to
\begin{equation}
  \label{eq:processfidelitypure}
  \mathcal{F}=\int_TdT \tr \left(\S_{\text{id}}(\varrho) \S(\varrho)\right),
  \quad
  (\varrho\in T),
\end{equation}
as $\S_{\text{id}}(\varrho)$ is a one-dimensional projector.
Throughout this paper we shall treat the latter case only, as we are
interested in cloning of pure states, which ideally results in
products of pure states.

According to Ref.~\cite{koenraad}, this optimization can be performed
in the following way. We consider a fixed basis on both the input and
output Hilbert spaces. $\S$ is sought for in its Choi-representation,
in which it is represented by a Hermitian, positive semidefinite
operator $X$ acting on $\mathcal{H}_\text{in}\otimes
\mathcal{H}_\text{out}$, and the relation of the output state to that
of the input reads
\begin{equation}
  \label{eq:choi}
  \S(\varrho) = \tr_{\mathcal{H}_\text{in}}
    \left( \left( 
      \varrho^T\otimes \hat 1 
      \right) X \right),
\end{equation}
where $T$ means ordinary (not Hermitian) transpose in the fixed basis,
and $\hat 1$ stands for the identity operator. In this representation
the process fidelity in Eq.~\eqref{eq:processfidelity} of the process
can be expressed as
\begin{equation}
  \label{eq:fidelityR}
  \mathcal{F}=\tr \left( X R\right),
\end{equation}
where 
\begin{equation}
  \label{eq:Rdef}
  R=\int_TdT \varrho^T \otimes \S_{\text{id}}(\varrho).
\end{equation}
The input of the optimization is the matrix $R$ encoding all the
relevant information on the set $T$ and the process $\S$ to be
approximated. The output will be the Choi matrix $X$ of the ideal
process and the maximum value $\mathcal{F}^\ast$ of the process
fidelity. Again note that for Eq.~\eqref{eq:fidelityR} to hold,
$\S_{\text{id}}(\varrho)$ has to be a pure state, that is, a
one-dimensional projector.

In the next step we fix two orthonormal bases: $\sigma$ and $\tau$, in
the linear space of the Hermitian matrices over
$\mathcal{H}_\text{in}$ and $\mathcal{H}_\text{out}$
respectively. Assuming $\dim \mathcal{H}_\text{in}=\dim
\mathcal{H}_\text{out}=d$, we have $d^2$ basis elements. We chose
$\sigma_0$ and $\tau_0$ to be proportional to the identity matrix,
whereas the other basis elements, indexed with positive integers,
should be traceless. For $d=2$ the Pauli matrices, for $d=3$ the
Gell-Mann matrices, while for higher dimensions, generalized Pauli
matrices are a suitable choice.

As derived in Ref.~\cite{koenraad}, one can construct the following
semidefinite program (in dual form, that is, in the form of matrix
inequalities) to optimize $\mathcal F$ in
Eq.~\eqref{eq:processfidelitypure}:
\begin{eqnarray}
  \label{eq:sdp}
  \text{maximize}\   &p = - c^T x \nonumber \\
  \text{subject to}\ &F_0+\sum_{\tilde{i}} x_{\tilde{i}}F_{\tilde{i}} \ge 0,
\end{eqnarray}
where $\ge0$ means positive semidefiniteness, $F_0=\frac1d\hat 1 $, and
\begin{equation}
  \label{eq:Fi}
  F_{\tilde{i}}=\sigma_{j(\tilde{i})}\otimes \tau_{k(\tilde{i})},
\end{equation}
and the indices $1\leq \tilde{i}\leq d^2(d^2-1)$ are chosen so that
the $j(\tilde{i})$-s take all the possible values from $0$ to $d^2$,
while the $k(\tilde{i})$-s take all the possible values from $1$ to
$d^2$, and the relation of the possible $(j,k)$ pairs and
$\tilde{i}$-s is bijective. The vector $x$ is the one to be found
while the constant coefficients $c$ in Eq.~\eqref{eq:sdp} are
defined as 
\begin{equation}
  \label{eq:c}
  c_{\tilde{i}}=-\tr \left(
    R F_{\tilde{i}} 
\right).
\end{equation}
Note that these coefficients encode the matrix $R$ of
Eq.~\eqref{eq:Rdef}, which, on the other hand, encodes all the
information about the process $\S_{\text{id}}$ to be approximated as
well as on the target set $T$. Having found the optimum $p^\ast$ of the
semidefinite program in Eq.~\eqref{eq:sdp}, the optimal fidelity is given
by
\begin{equation}
  \label{eq:Fopt}
  F^{\ast}= p + \frac 1d,
\end{equation}
whereas the Choi matrix of the process realizing it reads, in terms of
the vector $x$ corresponding to the optimum:
\begin{equation}
  \label{eq:choiopt}
 X_\text{opt}=\sum_{\tilde{i}}x_{\tilde{i}}F_{\tilde{i}}+\frac1d\hat 1.
\end{equation}
This is the recipe derived in Ref.~\cite{koenraad} for finding optimal
approximate realizations of quantum processes. It can be used
numerically to design approximate processes. As it requires a
semidefinite solver which is capable of handling Hermitian (complex)
matrices, SeDuMi~\cite{S98guide} appears to be a proper choice. To
invoke it for a semidefinite program formulated in Eq.~\eqref{eq:sdp},
it is very convenient to use the Matlab package of
T. Cubitt~\cite{quantinf}, which we have done in order to develop our
Matlab code producing the results described in what follows.

\subsection{Application for quantum cloning}
\label{sec:pocl}

Consider the case of quantum cloning. In general we are given $m$
copies of a physical system, all in the same identical quantum state,
say $|\Psi\rangle$. We would like to obtain $n>m$ systems, so that the
state of each system is closest in fidelity to $|\Psi\rangle$.  In the
simplest case, discussed in what follows, that of the $1\to 2$ qubit
cloners, we have a single qubit in state $|\Psi\rangle$, and we obtain two
copies in states $\varrho_1$ and $\varrho_2$ so that their fidelity
with respect to $|\Psi\rangle$ is maximal. If we want a symmetric cloner, for
which both fidelities are maximal, we have a problem characterized by
two objective functions. This is not suitable for the recipe described
in the previous section. Even if we consider the fidelity of the two
clones to be equal, it is not the kind of problem solved above with
semidefinite programming.

What we might consider instead is the following. Take the system in
state $|\Psi\rangle$, and an ancilla, in an arbitrary state, say
$|0\rangle$. The target set $T$ shall be the set for which we are
considering to plan a cloner, e.g.\ for a universal qubit cloner, the
surface of the Bloch-sphere, but we may consider restricted input
sets. Then carry out the procedure so that the ideal process should be
$|\Psi\rangle\otimes|0\rangle \to |\Psi\rangle\otimes |\Psi\rangle$, which
is obviously unfeasible due to the no-cloning theorem. The fidelity
$\mathcal{F}$ considered in this optimization is, however, not the
same as optimizing the fidelity of the clones. 

One may realize that it might be even a different
problem. Namely, we do not look for a process which produces two
identical copies which resemble the original to the highest extent,
but we look for a process which produces \emph{the product of the two
 originals} to the highest extent. For a symmetric $1\to 2$ cloner,
the fidelity of each clone to the original is considered, which will
be termed as \emph{cloning fidelity} in what follows. The cloners we
design here, on the other hand, are optimal with respect to the
process fidelity, hence, we shall call them \emph{process-optimized
  cloners}. 

%There are two questions addressed here, at least for the
%simplest, $1\to 2$ qubit case. Are the process optimized cloners
%optimal also with respect to cloning fidelity? Can a cloner be optimal
%with respect to cloning fidelity while being suboptimal with respect
%to process fidelity? 

Process-optimized cloners, as it was mentioned in the introduction,
were studied in the early literature of cloning. As it appears that
this kind of optimization is suitable for SDP, it is still of some
interest to study the actual designs which SDP provides automatically.

\section{Process-optimized quantum cloner designs}
\label{sec:cloners}

In this Section we describe our results regarding process-optimized
$1\to 2$ qubit cloners designed the above-described method. The
numerical results are summarized in Table~\ref{tab:cloners}.

\begin{table*}
\begin{tabular}{|p{8cm}||c|c|c|c|c|}
  \hline
  Cloner&$F_C$&$F_P$&$C$&$H_{\text{clone}}$&$H_{\text{out}}$
  \\
  \hline
  \hline
  UCQC, symmetric mode &
  $\frac56\approx 0.83333$ &
  $\frac23\approx 0.66667$ &
  $\frac13\approx 0.33333$ &
  $0.65002$ &
  $0.91830$ \\
\hline
universal, opt. for ancilla: $|0\rangle$, used ancilla: $|0\rangle$&0.83333&0.66667&0.33333&0.65002&0.91830\\ 
 \hline 
universal, opt. for ancilla: $\frac12\hat 1$, used ancilla: $\frac12\hat 1$&0.83333&0.66667&0.33333&0.65002&0.91830\\ 
 \hline 
universal, opt. for ancilla: $|0\rangle$, used ancilla: $\frac12\hat 1$&0.66667&0.45833&0.00000&0.91830&1.78434\\ 
 \hline 
{\bf universal, opt. for ancilla: $\frac12\hat 1$, used
  ancilla:$|0\rangle$} &{\bf 0.83333}&0.66667&0.33333&0.65002&0.91830\\ 
 \hline 
equator, opt. for ancilla: $|0\rangle$, used ancilla:
$|0\rangle$&0.83333&{\bf 0.75000}&0.33333&0.65002&0.65002\\ 
 \hline 
equator, opt. for ancilla: $\frac12\hat 1$, used ancilla: $\frac12\hat
1$&0.83333&{\bf 0.75000}&0.33333&0.65002&0.65002\\ 
 \hline 
equator, opt. for ancilla: $|0\rangle$, used ancilla: $\frac12\hat 1$&0.66667&0.50000&0.00000&0.91830&1.70058\\ 
 \hline 
{\bf equator, opt. for ancilla: $\frac12\hat 1$, used ancilla:
$|0\rangle$}&{\bf 0.83333}&{\bf 0.75000}&0.33333&0.65002&0.65002\\ 
 \hline 
\end{tabular}
\caption{Input-state independent parameters of the designed
  cloners. The most relevant issues are typeset in bold.}
\label{tab:cloners}
\end{table*} 

\subsection{The universal covariant quantum cloner}

First we calculate the properties of the qubit-version of the
universal covariant quantum cloner (UCQC) designed by Braunstein et
al.~\cite{PhysRevA.63.052313} We use this particular cloner design as
a benchmark for those designed by us. We use this cloner since it is
known to be optimal and it is a particular circuit so all of its
properties can be investigated.

\begin{figure}
  \centering
  \includegraphics[width=0.4\textwidth]{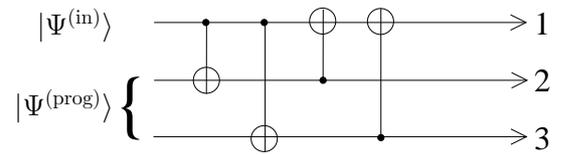}
  \caption{The quantum logic network for the universal quantum cloner
    of Ref.~\cite{PhysRevA.63.052313}, used here as a benchmark. Input
    ports are to to left, whereas the outputs to the right.}
  \label{fig:UCQC}
\end{figure}

The quantum logic network for the UCQC is depicted in
Fig.~\ref{fig:UCQC}.
The ``circuitry'' consists of four controlled NOT gates. We do not
consider all of its capabilities here, just one very particular case:
If a qubit in a quantum state $|\Psi_{\text{in}}\rangle$ to be cloned
impinges on port 1, while the so-called \emph{program state} is
chosen to be
\begin{equation}
  \label{eq:psiprog}
  |\Psi^{\text{prog}}\rangle = \frac1{\sqrt{6}} \left(
    2 |0\rangle|0\rangle+
    |0\rangle|1\rangle+
    |1\rangle|1\rangle \right),
\end{equation}
then in the outputs 1 and 2 there will be two identical clones of the
state to be cloned. (Port 3 will hold a state related to the input, it
will be omitted). Moreover, their density matrix will be the mixture
of the state to be cloned and a complete mixture. This case (i.e.\ the
symmetrical mode of this cloner) shall be referred to as UCQC in what
follows.

An obvious quantity to consider is the cloning fidelity, defined as
\begin{equation}
  \label{eq:cloningfidel}
  F_{\text{C}}=\langle\Psi_{\text{in}}|\varrho^{(1)}|\Psi_{\text{in}}\rangle,
\end{equation}
where $\varrho^{(1)}$ stands for the state of the first clone. (The
same can be calculated for the second clone, but as we consider
symmetric cloners here, these will be equal.). The UCQC attains the
optimal value of $5/6$, regardless of the state.
Another quantity introduced qualitatively in Section~\ref{sec:pocl}
reads
\begin{equation}
  \label{eq:processfidel}
    F_{\text{P}}=\left(\langle\Psi_{\text{in}}|\otimes\langle\Psi_{\text{in}}|\right)
      \varrho^{(12)}\left(|\Psi_{\text{in}}\rangle|\Psi_{\text{in}}\rangle\right),
\end{equation}
the fidelity of the joint state of the two clones with respect to the
product of two originals. For the UCQC it can be calculated, and it
will be $2/3$, regardless of the input state.

Further quantities of interest may be the entanglement as measured by
concurrence of the clones or the von Neumann entropy
\begin{equation}
  \label{eq:Entropy} H(\varrho)=-\tr \varrho \log_2 \varrho
\end{equation} of the clones and that of the two clones together.  The
concurrence is calculated according to the well-known Wootters
formula~\cite{PhysRevLett.80.2245}:
\begin{equation}
  \label{eq:Concurrence}
C(\varrho^{(12)})=\max(0,\lambda_1-\lambda_2-\lambda_3-\lambda_4),
\end{equation} the $\lambda$-s being the eigenvalues of the matrix
$\sqrt{\sqrt{\varrho^{(12)}}\tilde\varrho^{(12)}\sqrt{\varrho^{(12)}}}$
in descending order, whereas $\tilde
\varrho^{(12)}=\sigma_y\otimes\sigma_y
\varrho^{(12)T}\sigma_y\otimes\sigma_y$, $\sigma_y$ being the second
Pauli matrix.  Having evaluated these quantities for the UCQC, we have
listed their values are summarized in the first row of
Table~\ref{tab:cloners}.

\subsection{Universal cloners}

First we consider the state to be cloned an arbitrary one on the Bloch-sphere:
\begin{equation}
  \label{eq:Hbloch}
  |\Psi_{\text{in}}(\theta,\phi)\rangle =  \cos\left(\frac\theta2\right) |0\rangle
  +\sin\left(\frac\theta2\right) e^{i\phi}|1\rangle,
\end{equation}
hence, the cloner to be designed is universal. We apply the method
described in Section~\ref{sec:method}. Assume that we have an ancilla
initially in the state $\varrho_{\text{anc}}$. So the whole two-qubit state
impingement on the apparatus reads
\begin{equation}
  \label{eq:psiinuniv}
  \varrho_{\text{in}}(\theta,\phi)=|\Psi_{\text{in}}(\theta,\phi)\rangle
  \langle\Psi_{\text{in}}(\theta,\phi)|\otimes \varrho_{\text{anc}},
\end{equation}
and the output for an ideal cloner would be 
\begin{equation}
  \label{eq:psioutuniv}
  |\Psi_{\text{out,id}}(\theta,\phi)\rangle=
  |\Psi_{\text{in}}(\theta,\phi)\rangle\otimes|\Psi_{\text{in}}(\theta,\phi)\rangle
\end{equation}
This is to be substituted to the matrix $R$ of Eq.~\eqref{eq:Rdef}, yielding
\begin{widetext}
  \begin{equation}
    \label{eq:Runiv}
    R_{\text{univ}}=
    \frac{1}{4\pi}
    \int_0^\pi \sin(\theta) d\theta\,
    \int_0^{2\pi} d\phi\,
    \varrho_{\text{in}}^T(\theta,\phi)
    \otimes
    |\Psi_{\text{out,id}}(\theta,\phi)\rangle
        \langle \Psi_{\text{out,id}}(\theta,\phi)|,
  \end{equation}
  where we average over the surface of the Bloch-sphere. 

  The initial state of the ancilla is also an input for the
  optimization, providing another input to the problem. We considered
  two possibilities. If the ancilla is in the state $|0\rangle$, the
  matrix reads
  \begin{equation}
    \label{eq:Runivpureanc}
    R_{\text{univ,pure}}=
    \frac{1}{12}
    \begin{pmatrix}    
      3&0&0&0&0&0&0&0&0&1&1&0&0&0&0&0\\ 
      0&1&1&0&0&0&0&0&0&0&0&1&0&0&0&0\\ 
      0&1&1&0&0&0&0&0&0&0&0&1&0&0&0&0\\ 
      0&0&0&1&0&0&0&0&0&0&0&0&0&0&0&0\\ 
      0&0&0&0&0&0&0&0&0&0&0&0&0&0&0&0\\ 
      0&0&0&0&0&0&0&0&0&0&0&0&0&0&0&0\\ 
      0&0&0&0&0&0&0&0&0&0&0&0&0&0&0&0\\ 
      0&0&0&0&0&0&0&0&0&0&0&0&0&0&0&0\\ 
      0&0&0&0&0&0&0&0&1&0&0&0&0&0&0&0\\ 
      1&0&0&0&0&0&0&0&0&1&1&0&0&0&0&0\\ 
      1&0&0&0&0&0&0&0&0&1&1&0&0&0&0&0\\ 
      0&1&1&0&0&0&0&0&0&0&0&3&0&0&0&0\\ 
      0&0&0&0&0&0&0&0&0&0&0&0&0&0&0&0\\ 
      0&0&0&0&0&0&0&0&0&0&0&0&0&0&0&0\\ 
      0&0&0&0&0&0&0&0&0&0&0&0&0&0&0&0\\ 
      0&0&0&0&0&0&0&0&0&0&0&0&0&0&0&0
    \end{pmatrix},
  \end{equation}
while for the complete mixture as an ancilla we have
\begin{equation}
R_{\text{univ,cmix}}=
\frac{1}{24}
  \begin{pmatrix}
3&0&0&0&0&0&0&0&0&1&1&0&0&0&0&0\\
0&1&1&0&0&0&0&0&0&0&0&1&0&0&0&0
\\0&1&1&0&0&0&0&0&0&0&0&1&0&0&0&0
\\0&0&0&1&0&0&0&0&0&0&0&0&0&0&0&0
\\0&0&0&0&3&0&0&0&0&0&0&0&0&1&1&0
\\0&0&0&0&0&1&1&0&0&0&0&0&0&0&0&1
\\0&0&0&0&0&1&1&0&0&0&0&0&0&0&0&1
\\0&0&0&0&0&0&0&1&0&0&0&0&0&0&0&0
\\0&0&0&0&0&0&0&0&1&0&0&0&0&0&0&0
\\1&0&0&0&0&0&0&0&0&1&1&0&0&0&0&0
\\1&0&0&0&0&0&0&0&0&1&1&0&0&0&0&0
\\0&1&1&0&0&0&0&0&0&0&0&3&0&0&0&0
\\0&0&0&0&0&0&0&0&0&0&0&0&1&0&0&0
\\0&0&0&0&1&0&0&0&0&0&0&0&0&1&1&0
\\0&0&0&0&1&0&0&0&0&0&0&0&0&1&1&0
\\0&0&0&0&0&1&1&0&0&0&0&0&0&0&0&3
  \end{pmatrix}
\end{equation}
\end{widetext}
In spite of the sparsity of the matrices, we list all their elements,
since their structure is more visible. For obvious symmetry reasons,
one might consider any other pure state instead of $|0\rangle$, the
results would be equivalent.

Carrying out the optimization, we obtain the CP map given in
Appendix~\ref{clonerchoi}.
Evaluating the cloning behavior we found that in case of both of these
cloners, all the examined parameters are input-state
independent, thus we have designed a state-independent cloner, though
there was no constraint in the optimization to warrant this. 
The parameters of the designed cloners are listed in the
first four rows of Table~\ref{tab:cloners} for different
scenarios: we have applied both designs with both of the considered
ancillae.

As the most important consequence, it appears that for a universal
cloner, the process-optimized one's parameters are equal to those of
the  UCQC. Hence, this method designs an optimal
universal symmetric $1\to 2$ cloner. The optimality in terms of
process fidelity and that of cloning fidelity coincide. In addition,
the cloner is state-independent, though we have not prescribed
that. We have just optimized for average process fidelity.

Another consequence is that it is possible to design a cloner which
works for a completely mixed ancilla as well. Of course, we seek for
any CP maps, their realization may require several additional
ancillae, hence, one may combine the cloner designed for a pure
ancilla to a process which replaces complete mixture with a pure
ancilla. So this consequence is maybe less surprising. Nevertheless,
as one would expect according to the previous argument, the cloners
designed for the complete mixture as an ancilla perform optimally for
the pure input state as well, while those designed for the pure
ancilla operate as classical copiers (cloning fidelity of $\frac23$),
though the clones are unentangled.

It is interesting though to take a look at the Choi matrices of the
above-mentioned two cloners, given in Eqs.~\eqref{eq:choipureuniv}
and~\eqref{eq:Xunivanccmix} in the Appendix. Interestingly, the rank
of the Choi matrix (thus the number of Kraus operators in the
orthogonal Kraus representation) is $4$ for the completely mixed
ancilla, while it is $10$ for the pure one. This suggests that the
cloner for the mixed ancilla is a ``simpler'' operation than that for
the pure one.

In summary, in case of an $1\to 2$ universal symmetric qubit cloner,
the optimization of process fidelity yields a state-independent cloner
which performs, at least in terms of the examined parameters, exactly
as the UCQC.

\subsection{Phase-covariant cloners}

Now we restrict our attention to the optimization for the equator of
the Bloch-sphere, that is, phase-covariant cloning. In this case the
matrix $R$ in Eq.~\eqref{eq:Runiv}, the integral becomes a line
integral:
\begin{widetext}
  \begin{equation}
    \label{eq:Requator}
    R_{\text{equator}}=
    \frac{1}{2\pi}
    \int_0^{2\pi} d\phi\,
    \varrho_{\text{in}}^T(\theta=\frac{\pi}{2},\phi)
    \otimes
    |\Psi_{\text{out,id}}(\theta=\frac{\pi}{2},\phi)\rangle
        \langle \Psi_{\text{out,id}}(\theta=\frac{\pi}{2},\phi)|,
  \end{equation}
and the matrix will read
  \begin{equation}
    \label{eq:Requatorn}
    R_{\text{equator}}=
    \frac18
    \begin{pmatrix}
      1&0&0&0&0&0&0&0&0&1&1&0&0&0&0&0
\\0&1&1&0&0&0&0&0&0&0&0&1&0&0&0&0
\\0&1&1&0&0&0&0&0&0&0&0&1&0&0&0&0
\\0&0&0&1&0&0&0&0&0&0&0&0&0&0&0&0
\\0&0&0&0&0&0&0&0&0&0&0&0&0&0&0&0
\\0&0&0&0&0&0&0&0&0&0&0&0&0&0&0&0
\\0&0&0&0&0&0&0&0&0&0&0&0&0&0&0&0
\\0&0&0&0&0&0&0&0&0&0&0&0&0&0&0&0
\\0&0&0&0&0&0&0&0&1&0&0&0&0&0&0&0
\\1&0&0&0&0&0&0&0&0&1&1&0&0&0&0&0
\\1&0&0&0&0&0&0&0&0&1&1&0&0&0&0&0
\\0&1&1&0&0&0&0&0&0&0&0&1&0&0&0&0
\\0&0&0&0&0&0&0&0&0&0&0&0&0&0&0&0
\\0&0&0&0&0&0&0&0&0&0&0&0&0&0&0&0
\\0&0&0&0&0&0&0&0&0&0&0&0&0&0&0&0
\\0&0&0&0&0&0&0&0&0&0&0&0&0&0&0&0
\end{pmatrix}.
  \end{equation}
\end{widetext}
Note that there is very little difference between this matrix and that
of the universal cloner with pure ancilla in
Eq.~\eqref{eq:Runivpureanc}. 

Carrying out the optimization we obtain the cloner given exactly in
Appendix~\ref{clonerchoi}.  Importantly, for the states on the equator
this cloner is also state-independent. Its data are listed in the
fifth row of Table~\ref{tab:cloners}. What is important to note that
almost all parameters are equal to those of the UCQC, \emph{except for
  the purity of the two clones together and the process fidelity}.
For the equator of the Bloch sphere, the process optimized one, being
still an optimal cloner, attains a higher process fidelity. The value
of $0.75$ coincides with the one calculated analytically in
Ref.~\cite{DAriano}.

\begin{figure*}
    \centering
    \includegraphics[width=0.7\textwidth]{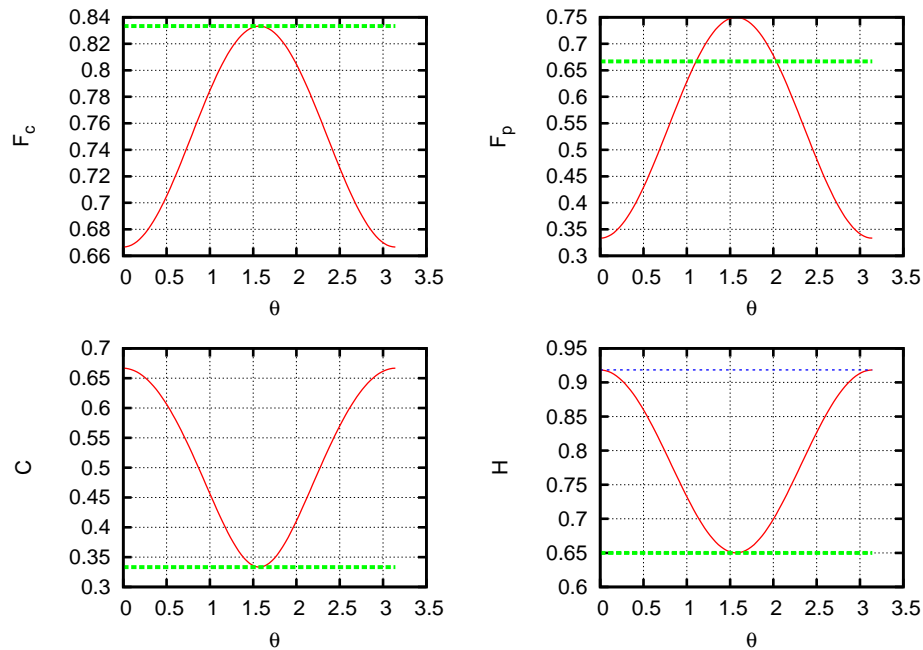}
    \caption{(color online) Quantities of the equatorial cloner as a
      function of the angle $\theta$ on the Bloch
      sphere. $F_c$ is the cloning fidelity, $F_p$ is the process
      fidelity, $C$ is the concurrence of the clones, $H$ is the von
      Neumann entropy. In all the figures, the respective quantity of
      the UCQC is also plotted, these are the
      straight lines, the curves represent the quantity for the
      equatorial cloner. In case of the entropy the entropy of the
      clone is always equal to the entropy of the bipartite states of
      the two clones. The lower straight line is the entropy of the
      clone, while the upper is the entropy of the bipartite state for
      the UCQC.}
    \label{fig:equator2D}
  \end{figure*}

It is also worth analyzing the behavior of the cloner designed for the
equator on the rest of the Bloch-sphere. For symmetry reasons it is
sufficient to investigate a meridian of the sphere, that is, the
dependence of parameters on $\theta $ with, say, $\phi =0$. These
functions are plotted in Fig.~\ref{fig:equator2D}, where the
parameters for the UCQC are also plotted for reference. It appears
that the parameters reach the values of the UCQC for
$\theta=\frac\pi2$, that is, the equator. As for the behavior of the
von Neumann entropy, the entropy of each clone is equal to that of the
system of two clones. In the optimal case this reaches the entropy of
clones of the UCQC, while at the ``poles'' of the sphere it reaches
the value of the joint two-clone system of the UCQC. It should be
noted also that while the cloning fidelity reaches $5/6$ only at the
equator, the process fidelity is superior to that of the UCQC for a
broader range of $\theta$-s.

We have also carried out the same analysis with the completely mixed
ancilla as for the case of the universal cloner. As reflected by the
last four lines of Table~\ref{tab:cloners}, the conclusion to be drawn is
the same as for the universal case: the cloner for pure ancilla
becomes a classical copier for a completely mixed ancilla. It is
possible to design a cloner for the completely mixture, which works
for the pure ancilla as well. 

Again it is interesting though to take a look at the Choi matrices of the
above cloners, given in Eqs.~\eqref{eq:choiequatorpure}
and~\eqref{eq:Xequatorcmix}. It appears that the spectrum of these
matrices is the same as that of the universal cloners.

\begin{figure}
  \centering
\includegraphics[width=0.4\textwidth]{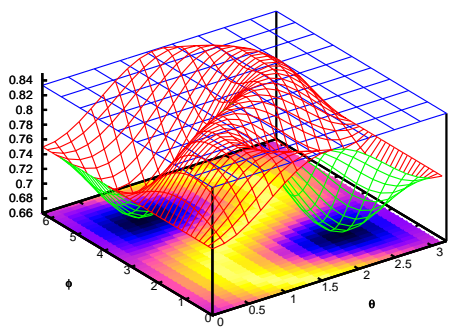}    
\caption{(color online) The cloning fidelity as a function of the
  spherical coordinates on the Bloch sphere of the input states, for a
  cloner designed for cloning the equator of the Bloch sphere rotated by
  $\pi/4$ about the $x$ axis. The upper flat plane represents $5/6$,
  the fidelity of the optimal cloner. The plotted quantity is
  dimensionless.}
  \label{fig:3dequator}
\end{figure}

Another question to be addressed in the case of a phase-covariant
cloner is that of the dependence on the ancilla. To investigate this
we have considered a case when the target set is the equator rotated
by a given angle about the $x$ axis. Carrying out this analysis for
various angles, we have found that the resulting cloner has the very
parameters of the above-detailed equatorial cloner: the designed
cloner is optimal and anisotropic, independently of the angle between
the ancilla state and the chosen main circle.  This is illustrated in
terms of cloning fidelity in Fig.~\ref{fig:3dequator}, for a main
circle rotated by $\pi/4$. This is not very surprising for symmetry
reasons, nevertheless it again demonstrates the power of the applied
numerical technique.

\section{Conclusions and outlook}
\label{sec:conclusion}

We have explicitly designed universal and phase covariant process
optimized quantum cloners: quantum cloners that are optimized for the
fidelity of the state of the whole set of output states to that of a
product of ideal clones. We studied their design for $1\to 2$ qubit
cloners numerically, via semidefinite programming. In all the studied
cases the so-designed cloners have been found to be optimal cloners
with respect to the fidelity of clones to the originals, too. 
% However,
% in the case of restricted-input cloners, the so-designed ones provide
% an output state of two optimal clones which are, together, closer in
% fidelity to a product of two ideal clones.

For the phase covariant case we have found the result to be expected
according to Ref.~\cite{DAriano}. We have analyzed the operation of
this cloner on the whole Bloch sphere as well, with respect to state
purity and entanglement.

% The results naturally raise the question whether the process-optimized
% cloners are better for certain quantum information tasks, such as,
% e.g.\ eavesdropping in quantum cryptography, than those designed for
% optimal cloning. Also, it would be interesting to examine the relation
% of our cloners to other particular results, and generalize the
% consideration to higher dimensions and larger number of clones. The
% framework provided here is obviously suitable for such a
% study. Finally, we believe that this framework, if considered
% analytically in more detail, could provide a deeper understanding of
% the no-cloning theorem and quantum cloning.

Our result demonstrates the power of the method introduced in
Ref.~\cite{koenraad} in designing particular quantum circuits for
different purposes in a neat and automated numerical way. The
only requirement for the process to be approximated is that its ideal
output should be pure states and one has to be able to average over
the possible input states.

\acknowledgements
  M.~K. and L.~D. acknowledge the support of the grant
  ``SROP-4.2.2/B-10/1-2010-0029 Supporting Scientific Training of
  Talented Youth at the University of P\'ecs''. M.~K. acknowledges the
  support of the The Hungarian Scientific Research Fund (OTKA) under
  the contract No. K83858. The authors thank J\'anos Bergou for
  inspiring ideas and useful discussions as well as Toma\v s Ryb\'ar for
  very important notes regarding the scientific context of the present
  paper.

%\bibliographystyle{prsty}
%\bibliography{bibliography_new}

%\newpage
\appendix
\widetext
\section{Choi matrices of some cloners}
\label{clonerchoi}

In what follows we write rational numbers as matrix elements. The
results of our calculations were their floating point counterparts
within numerical precision, so we use the rational counterparts for
sake of clarity. It would be possible to verify analytically that
these are the optima indeed, by calculating the the duality
gap in the optimization.

\paragraph{Universal cloner with pure ancilla}
The nonzero elements of the Choi matrix are:
\begin{equation}
  \label{eq:choipureuniv}
  X=
    \begin{pmatrix}    
       \frac23 &0&0&0&0&0&0&0&0&\frac13 &\frac13 &0&0&0&0&0\\ 
      0&\frac16 &\frac16 &0&0&0&0&0&0&0&0&\frac13 &0&0&0&0\\ 
      0&\frac16 &\frac16 &0&0&0&0&0&0&0&0&\frac13 &0&0&0&0\\ 
      0&0&0&0&0&0&0&0&0&0&0&0&0&0&0&0\\ 
      0&0&0&0&\frac14 &0&0&0&0&0&0&0&0&0&0&0\\ 
      0&0&0&0&0&\frac14 &0&0&0&0&0&0&0&0&0&0\\ 
      0&0&0&0&0&0&\frac14 &0&0&0&0&0&0&0&0&0\\ 
      0&0&0&0&0&0&0&\frac14 &0&0&0&0&0&0&0&0\\ 
      0&0&0&0&0&0&0&0&0&0&0&0&0&0&0&0\\ 
      \frac13 &0&0&0&0&0&0&0&0&\frac16 &\frac16 &0&0&0&0&0\\ 
      \frac13 &0&0&0&0&0&0&0&0&\frac16 &\frac16 &0&0&0&0&0\\ 
      0&\frac13 &\frac13&0&0&0&0&0&0&0&0&\frac23 &0&0&0&0\\ 
      0&0&0 &0&0&0&0&0&0&0&0&0&\frac14 &0&0&0\\ 
      0&0&0&0&0&0&0&0&0&0&0&0&0&\frac14 &0&0\\ 
      0&0&0&0&0&0&0&0&0&0&0&0&0&0&\frac14 &0\\ 
      0&0&0&0&0&0&0&0&0&0&0&0&0&0&0&\frac14
    \end{pmatrix}
\end{equation}
The rank of this matrix is 10. Its eigenvalues are $1.0$ (multiplicity
$2$) and $0.25$ (multiplicity $8$)
  %X_{1,1} = \frac23\nonumber \\
 %X_{10,1} = \frac13\nonumber \\
 %X_{11,1} = \frac13\nonumber \\
 % X_{2,2} = \frac16\nonumber \\
  %X_{3,2} = \frac16\nonumber \\
% X_{12,2} = \frac13\nonumber \\
  %X_{2,3} = \frac16\nonumber \\
 % X_{3,3} = \frac16\nonumber \\
 %X_{12,3} = \frac13\nonumber \\
  %X_{5,5} = \frac14\nonumber \\
  %X_{6,6} = \frac14\nonumber \\
  %X_{7,7} = \frac14\nonumber \\
  %X_{8,8} = \frac14\nonumber \\
 % X_{1,10}= \frac13\nonumber \\
 %X_{10,10}= \frac16\nonumber \\
% X_{11,10}= \frac16\nonumber \\
 % X_{1,11}= \frac13\nonumber \\
 %X_{10,11}= \frac16\nonumber \\
 %X_{11,11}= \frac16\nonumber \\
 % X_{2,12}= \frac13\nonumber \\
 % X_{3,12}= \frac13\nonumber \\
 %X_{12,12}= \frac23\nonumber \\
 %X_{13,13}= \frac14\nonumber \\
 %X_{14,14}= \frac14\nonumber \\
 %X_{15,15}= \frac14\nonumber \\
 %X_{16,16}= \frac14

\paragraph{Universal cloner with completely mixed ancilla}
The Choi matrix reads:
\begin{equation}
  \label{eq:Xunivanccmix}
   X=
    \begin{pmatrix}    
      \frac23 &0&0&0&0&0&0&0&0&\frac13 &\frac13 &0&0&0&0&0\\ 
      0&\frac16 &\frac16 &0&0&0&0&0&0&0&0&\frac13 &0&0&0&0\\ 
      0&\frac16 &\frac16 &0&0&0&0&0&0&0&0&\frac13 &0&0&0&0\\ 
      0&0&0&0&0&0&0&0&0&0&0&0&0&0&0&0\\ 
      0&0&0&0&\frac23 &0&0&0&0&0&0&0&0&\frac13 &\frac13 &0\\ 
      0&0&0&0&0&\frac16 &\frac16 &0&0&0&0&0&0&0&0&\frac13\\ 
      0&0&0&0&0&\frac16 &\frac16 &0&0&0&0&0&0&0&0&\frac13\\ 
      0&0&0&0&0&0&0&0&0&0&0&0&0&0&0&0\\ 
      0&0&0&0&0&0&0&0&0&0&0&0&0&0&0&0\\ 
      \frac13 &0&0&0&0&0&0&0&0&\frac16 &\frac16 &0&0&0&0&0\\ 
      \frac13 &0&0&0&0&0&0&0&0&\frac16 &\frac16 &0&0&0&0&0\\ 
      0&\frac13 &\frac13 &0&0&0&0&0&0&0&0&\frac23 &0&0&0&0\\ 
      0&0&0&0&0&0&0&0&0&0&0&0&0&0&0&0\\ 
      0&0&0&0&\frac13 &0&0&0&0&0&0&0&0&\frac16 &\frac16 &0\\ 
      0&0&0&0&\frac13 &0&0&0&0&0&0&0&0&\frac16 &\frac16 &0\\ 
      0&0&0&0&0&\frac13 &\frac13 &0&0&0&0&0&0&0&0&\frac23
    \end{pmatrix}
 % X_{1,1}   =\frac23\nonumber \\
% X_{10,1} = \frac13\nonumber \\
% X_{11,1} = \frac13\nonumber \\
  %X_{2,2} = \frac16\nonumber \\
 % X_{3,2} = \frac16\nonumber \\
% X_{12,2} = \frac13\nonumber \\
  %X_{2,3} = \frac16\nonumber \\
 % X_{3,3} = \frac16\nonumber \\
 %X_{12,3} = \frac13\nonumber \\
  %X_{5,5} = \frac23\nonumber \\
% X_{14,5} = \frac13\nonumber \\
% X_{15,5} = \frac13\nonumber \\
 % X_{6,6} = \frac16\nonumber \\
 % X_{7,6} = \frac16\nonumber \\
% X_{16,6} = \frac13\nonumber \\
 % X_{6,7} = \frac16\nonumber \\
  %X_{7,7} = \frac16\nonumber \\
% X_{16,7} = \frac13\nonumber \\
 % X_{1,10}= \frac13\nonumber \\
% X_{10,10}= \frac16\nonumber \\
% X_{11,10}= \frac16\nonumber \\
%  X_{1,11}= \frac13\nonumber \\
 %X_{10,11}= \frac16\nonumber \\
%X_{11,11}= \frac16\nonumber \\
 % X_{2,12}= \frac13\nonumber \\
 % X_{3,12}= \frac13\nonumber \\
% X_{12,12}= \frac23\nonumber \\
 % X_{5,14}= \frac13\nonumber \\
 %X_{14,14}= \frac16\nonumber \\
 %X_{15,14}= \frac16\nonumber \\
  %X_{5,15}= \frac13\nonumber \\
 %X_{14,15}= \frac16\nonumber \\
 %X_{15,15}= \frac16\nonumber \\
 % X_{6,16}= \frac13\nonumber \\
 % X_{7,16}= \frac13\nonumber \\
 %X_{16,16}= \frac23
\end{equation}
The rank of this matrix is $4$, the only eigenvalue is $1.0$.

\paragraph{Equator, pure ancilla}
The Choi matrix reads:
\begin{equation}
  \label{eq:choiequatorpure}
   X=
    \begin{pmatrix}    
       \frac13 &0&0&0&0&0&0&0&0& \frac13 & \frac13 &0&0&0&0&0\\ 
      0& \frac13 & \frac13 &0&0&0&0&0&0&0&0& \frac13 &0&0&0&0\\ 
      0& \frac13 & \frac13 &0&0&0&0&0&0&0&0& \frac13 &0&0&0&0\\ 
      0&0&0&0&0&0&0&0&0&0&0&0&0&0&0&0\\ 
      0&0&0&0&\frac14 &0&0&0&0&0&0&0&0&0&0&0\\ 
      0&0&0&0&0&\frac14 &0&0&0&0&0&0&0&0&0&0\\ 
      0&0&0&0&0&0&\frac14 &0&0&0&0&0&0&0&0&0\\ 
      0&0&0&0&0&0&0&\frac14 &0&0&0&0&0&0&0&0\\ 
      0&0&0&0&0&0&0&0&0&0&0&0&0&0&0&0\\ 
       \frac13 &0&0&0&0&0&0&0&0& \frac13 & \frac13 &0&0&0&0&0\\ 
       \frac13 &0&0&0&0&0&0&0&0& \frac13 & \frac13 &0&0&0&0&0\\ 
      0& \frac13 & \frac13 &0&0&0&0&0&0&0&0& \frac13 &0&0&0&0\\ 
      0&0&0&0&0&0&0&0&0&0&0&0&\frac14 &0&0&0\\ 
      0&0&0&0&0&0&0&0&0&0&0&0&0&\frac14 &0&0\\ 
      0&0&0&0&0&0&0&0&0&0&0&0&0&0&\frac14 &0\\ 
      0&0&0&0&0&0&0&0&0&0&0&0&0&0&0&\frac14
    \end{pmatrix}
%  X_{1,1}  = \frac13\nonumber \\
% X_{10,1} = \frac13\nonumber \\
% X_{11,1} = \frac13\nonumber \\
  %X_{2,2} = \frac13\nonumber \\
 % X_{3,2} = \frac13\nonumber \\
% X_{12,2} = \frac13\nonumber \\
 % X_{2,3} = \frac13\nonumber \\
 % X_{3,3} = \frac13\nonumber \\
 %X_{12,3} = \frac13\nonumber \\
 % X_{5,5} = \frac14\nonumber \\
 % X_{6,6} = \frac14\nonumber \\
 % X_{7,7} = \frac14\nonumber \\
 % X_{8,8} = \frac14\nonumber \\
 % X_{1,10}= \frac13\nonumber \\
 %X_{10,10}= \frac13\nonumber \\
% X_{11,10}= \frac13\nonumber \\
%  X_{1,11}= \frac13\nonumber \\
 %X_{10,11}= \frac13\nonumber \\
%X_{11,11}= \frac13\nonumber \\
 % X_{2,12}= \frac13\nonumber \\
 % X_{3,12}= \frac13\nonumber \\
% X_{12,12}= \frac13\nonumber \\
% X_{13,13}= \frac14\nonumber \\
% X_{14,14}= \frac14\nonumber \\
% X_{15,15}= \frac14\nonumber \\
% X_{16,16}= \frac14
\end{equation}
Its rank is $10$, with the same eigenvalues and multiplicities as the
matrix for the universal cloner with pure ancilla.

\paragraph{Equator, completely mixed ancilla}
The Choi matrix reads:
\begin{equation}
  \label{eq:Xequatorcmix}
   X=
  \begin{pmatrix}    
       \frac13 &0&0&0&0&0&0&0&0& \frac13 & \frac13 &0&0&0&0&0\\ 
      0& \frac13 & \frac13 &0&0&0&0&0&0&0&0& \frac13 &0&0&0&0\\ 
      0& \frac13 & \frac13 &0&0&0&0&0&0&0&0& \frac13 &0&0&0&0\\ 
      0&0&0&0&0&0&0&0&0&0&0&0&0&0&0&0\\ 
      0&0&0&0&\frac13 &0&0&0&0&0&0&0&0&\frac13 &\frac13 &0\\ 
      0&0&0&0&0&\frac13 &\frac13 &0&0&0&0&0&0&0&0&\frac13\\ 
      0&0&0&0&0&\frac13 &\frac13 &0&0&0&0&0&0&0&0&\frac13\\ 
      0&0&0&0&0&0&0&0&0&0&0&0&0&0&0&0\\ 
      0&0&0&0&0&0&0&0&0&0&0&0&0&0&0&0\\ 
       \frac13 &0&0&0&0&0&0&0&0& \frac13 & \frac13 &0&0&0&0&0\\ 
       \frac13 &0&0&0&0&0&0&0&0& \frac13 & \frac13 &0&0&0&0&0\\ 
      0& \frac13 & \frac13 &0&0&0&0&0&0&0&0& \frac13 &0&0&0&0\\ 
      0&0&0&0&0&0&0&0&0&0&0&0&0 &0&0&0\\ 
      0&0&0&0&\frac13 &0&0&0&0&0&0&0&0&\frac13 &\frac13 &0\\ 
      0&0&0&0&\frac13 &0&0&0&0&0&0&0&0&\frac13 &\frac13 &0\\ 
      0&0&0&0&0&\frac13 &\frac13 &0&0&0&0&0&0&0&0&\frac13
    \end{pmatrix}
 % X_{1,1} = \frac13\nonumber \\
 %X_{10,1} = \frac13\nonumber \\
 %X_{11,1} = \frac13\nonumber \\
 % X_{2,2} = \frac13\nonumber \\
 % X_{3,2} = \frac13\nonumber \\
 %X_{12,2} = \frac13\nonumber \\
%  X_{2,3} = \frac13\nonumber \\
%  X_{3,3} = \frac13\nonumber \\
 %X_{12,3} = \frac13\nonumber \\
 % X_{5,5} = \frac13\nonumber \\
 %X_{14,5} = \frac13\nonumber \\
 %X_{15,5} = \frac13\nonumber \\
  %X_{6,6} = \frac13\nonumber \\
  %X_{7,6} = \frac13\nonumber \\
% X_{16,6} = \frac13\nonumber \\
 % X_{6,7} = \frac13\nonumber \\
 % X_{7,7} = \frac13\nonumber \\
 %X_{16,7} = \frac13\nonumber \\
 % X_{1,10}= \frac13\nonumber \\
% X_{10,10}= \frac13\nonumber \\
 %X_{11,10}= \frac13\nonumber \\
 % X_{1,11}= \frac13\nonumber \\
% X_{10,11}= \frac13\nonumber \\
%X_{11,11}= \frac13\nonumber \\
 % X_{2,12}= \frac13\nonumber \\
 % X_{3,12}= \frac13\nonumber \\
% X_{12,12}= \frac13\nonumber \\
  %X_{5,14}= \frac13\nonumber \\
 %X_{14,14}= \frac13\nonumber \\
 %X_{15,14}= \frac13\nonumber \\
  %X_{5,15}= \frac13\nonumber \\
 %X_{14,15}= \frac13\nonumber \\
 %X_{15,15}= \frac13\nonumber \\
  %X_{6,16}= \frac13\nonumber \\
  %X_{7,16}= \frac13\nonumber \\
% X_{16,16}= \frac13
\end{equation}
Its rank is $4$, with the same eigenvalue and multiplicities as the
universal cloner with completely mixed ancilla.
%\end{widetext}

\end{document}